\documentclass[reprint,amsmath,amssymb,aps,]{revtex4-1}

\usepackage{graphicx}
\usepackage{dcolumn}
\usepackage{bm}
\usepackage{gensymb}
\usepackage{comment}

\begin{document}

\title{Features of heterogeneously charged systems at their liquid-liquid critical point}

\author{Daniele Notarmuzi}
\email{daniele.notarmuzi@tuwien.ac.at}
\affiliation{Institut für Theoretische Physik, TU Wien, Wiedner Hauptstraße 8-10, A-1040 Wien, Austria}
\author{Emanuela Bianchi}
\email{emanuela.bianchi@tuwien.ac.at}
\affiliation{Institut f\"{u}r Theoretische Physik, TU Wien, Wiedner Hauptstraße 8-10, A-1040 Wien, Austria and CNR-ISC, Uos Sapienza, Piazzale A. Moro 2, 00185 Roma, Italy}

\begin{abstract}
Recently synthesized colloids and biological systems such as proteins, viruses and monoclonal antibodies are heterogeneously charged, i.e.,  different regions of their surfaces carry different amount of positive or negative charge. Because of charge anisotropy, the electrostatics interactions between these units through the surrounding medium are intristically anisotropic, meaning that they are characterized not only by the attraction between oppositely-charged regions but also by
the repulsion between like-charged areas. Recent experiments have shown that the liquid-liquid phase separation of these systems can be driven by anisotropic electrostatics interactions, but it is not clear how  the emerging aggregates are affected by charge imbalance and charge patchiness. The ability to experimentally control these two quantities calls for a theoretical understanding of their interplay, which we address here at the critical point. We consider a coarse grained model of anisotropically charged hard spheres whose interaction potential is grounded in a robust mean field theory and perform extensive numerical Monte Carlo simulations to understand the aggregation behavior of these units at the critical point. Stemming from the simplicity of the model, we address the interplay between  charge imbalance and charge patchiness with the use of three parameters only and fully rationalize how these features impact the critical point of the model by means of thermodynamic-independent pair properties. 
\end{abstract}

\maketitle

\section{Introduction}
 
Colloidal particles featuring engineered surface patterns serve both as self-assembling units for crafting new materials with target structures and properties\cite{Granick_2011,Pine_2020,Liedl_2024,Sulc_2024}, and as simple models to shed light on the aggregation behaviors observed in biological systems, such as globular proteins, viral capsids and antibodies~\cite{cademartiri,Vacha_2017,Schurtenberger_2020,Espinoza2020Liquid,Stradner_2019}.

Particle models with a built-in directional attraction, often referred to as patchy colloids, 
have shown a plethora of diverse collective behaviours, such as the formation of finite clusters with well-defined geometries, the assembly of exotic two- and three-dimensional crystals and the emergence of disordered networks with incessantly rearranging topology, to name just a few examples~\cite{Williamson,Kretzschmar_2010,Bianchi_pccp_2011,Romano_2012,SmallenburgLeibler_2013,Bianchi2017Limiting,Chakrabarti_2018}.
In the context of the liquid-liquid phase separation (LLPS),  i.e., the separation into a dilute and a dense disordered phase, patchy colloid models have provided insight into the stability of the LLPS, with particular reference to globular proteins~\cite{Thomson1987binary,schurtenberger1989observation,chen2004liquid,annunziata2005oligomerization,Schurtenberger_2011}: when the particle bonding valence is limited (due to the built-in particle functionality), then the LLPS becomes metastable with respect to the liquid--crystal transition and a large region of the phase diagram becomes dominated by a homogeneous, low density liquid (often referred to as empty liquid) and, on gradually reducing the temperature, by a disordered arrested network (also referred to as ideal/equilibrium gel)~\cite{Sear_1999,KernFrenkel_2003,Bianchi2006Phase,Espinoza2020Liquid}. 

Particle models with directional repulsion on the top of the built-in directional attraction have been recently put forward to take into account the possibility of charge heterogeneity on particle surface~\cite{hoffmann2004molphys,boon2010jpcm,Bianchi2011Inverse,degraaf2012jcp,yigit2015jcp,hieronimus2016jcp,Brunk_2020,mathews2022molsim,popov2023jpcb} for charged patchy colloids~\cite{vanostrum2015jpcm,Virk2023Synthesis,sabapathy2017pccp,Lebdioua2021Study,Mehr2019sm} as well as globular proteins~\cite{Bozic2017pH,lund2016anisotropic,nakamura1985nature}. Models aiming at elucidating the role of charge patchiness have been used to investigate a variety of phenomena spanning from the bulk aggregation of charged Janus and patchy colloids~\cite{Cruz_2016,Vashisth_2021,Ferreira_2017} to the protein adsorption on polyelectrolyte chains or brush layers~\cite{Yigit15b,Yigit17}. 
The competition between attractive and repulsive charge–charge interactions has also been investigated in the context of the LLPS~\cite{Notarmuzi_2024,Blanco_2016}: in particular, we have recently shown that the interplay between the net particle charge and the surface patchiness controls the critical parameters of the LLPS in systems of model particles with a null dipole moment and a linear quadrupole moment~\cite{Notarmuzi_2024}. We consider here a broader and more systematic selection of systems with the aim of fully elucidating the trends of all thermodynamic parameters at the critical point on smoothly varying the surface anisotropy and the charge imbalance. 

The paper is organized as detailed in the following. In section~\ref{sec:model} we introduce the coarse-grained model, its microscopic background, features and parameters. In section~\ref{sec:methods}, we briefly discuss the details of our Monte Carlo simulations and the methods used to determine the critical points. In section~\ref{sec:results} we present our results. Namely in subsection~\ref{sec:criticalpoints} we discuss the critical parameters and fields on varying the interplay between directional repulsion and directional attraction as well as on changing the surface patchiness; we then relate the behaviour of the observed critical temperatures to (i) a thermodynamic-independent pair quantity that estimates the particles' availability to form bonds (subsection~~\ref{sec:VbTc}) and (ii) the reduced second virial coefficient at the estimated critical points (subsection~\ref{sec:b2}); moreover we relate the behaviour of the observed critical density to the morphology of the aggregates at the critical point  by (i) comparing the energy distributions of random versus simulated pairs of particles(subsection~\ref{sec:PProle}) and (ii) evaluating the number of bonds formed in the systems (subsection~\ref{sec:GbEb}).  We draw our conclusions  in section~\ref{sec:conclusions}
 
\section{The model}\label{sec:model}

\begin{figure*}[h]
\centering
  \includegraphics[width=\textwidth]{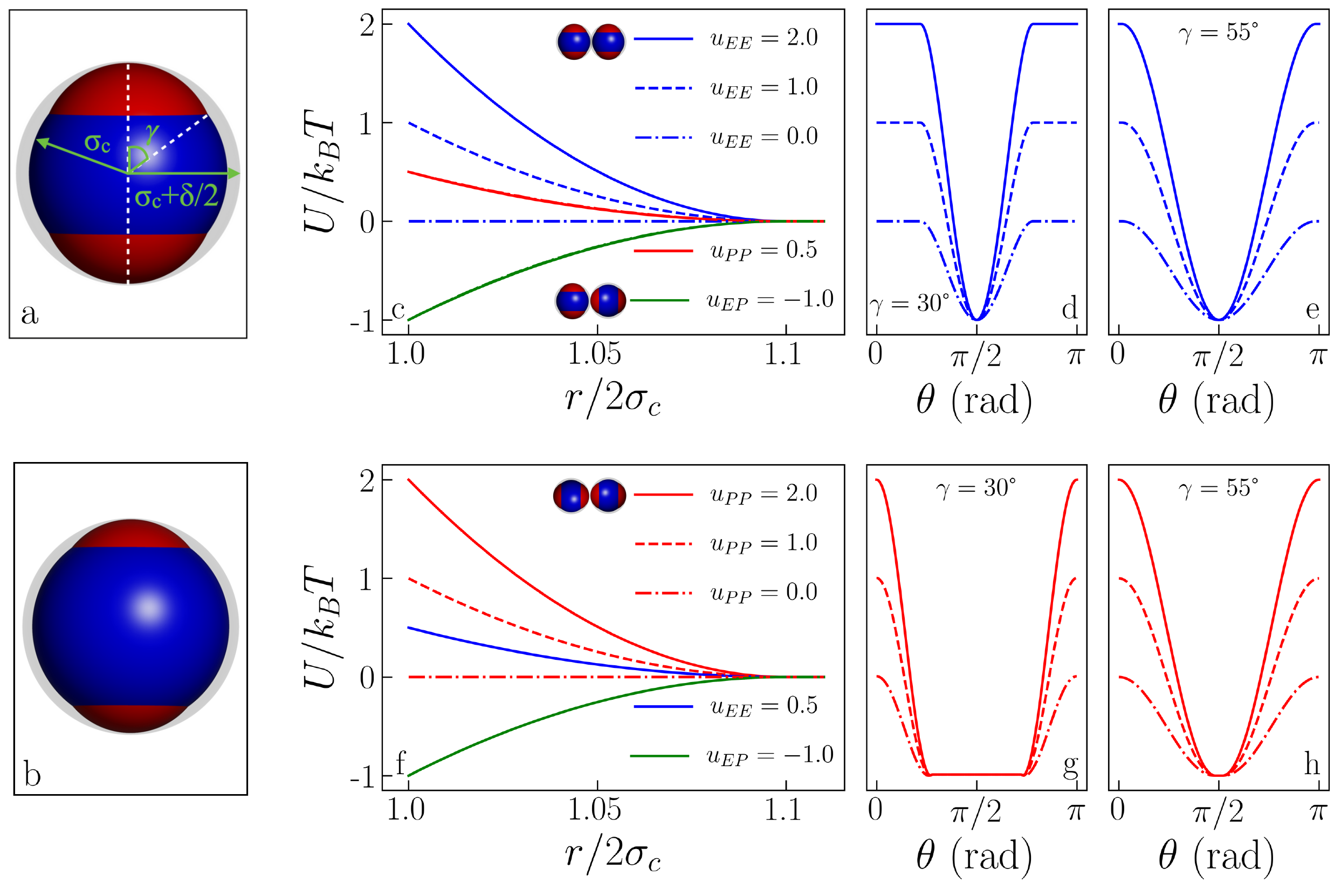}
\caption{Inverse Patchy Particle (IPP) model. (a) IPP particle sketch for $\gamma=55$°: the left green arrow 
represents the particle radius $\sigma_c=0.5$, the right green arrow 
represents the particle interaction radius $\sigma_c+\delta/2$, the white dashed vertical line represents the symmetry axis of the model which, together with the white dashed diagonal line, defines the half-opening angle $\gamma$, also shown in green, quantifying the patch extension. (b) IPP particle sketch for $\gamma=30^{\degree}$.
(c) Pair interaction energy as a function of the center-to-center distance for $\gamma=30^{\degree}$ and different energy sets: while  $u_{PP}=0.5$ is fixed (as much as $u_{EP}=-1.0$), $u_{EE}$ assumes the values $=2.0, 1.0, 0.0$, as labeled. The $EE$ reference configuration is shown by the upper pair of particles, while the lower pair of particles depicts the $EP$
reference configuration. (d) Pair interaction energy as the symmetry axis of one particle is rotated from $EE$ ($\theta=0$) to $EP$ ($\theta=\pi/2$) and back to $EE$ ($\theta=\pi$), for $\gamma=30^{\degree}$ and for the three $u_{EE}$ values in the legend of panel (c). (e) Same as in (d) but for $\gamma=55^{\degree}$. (f) Same as in (c) where $u_{PP}=2.0, 1.0, 0.0$ and $u_{EE}=0.5$. The reference $PP$ configuration is shown by the upper pair of particles.  (g) Same as in (d) but the starting configuration is $PP$ and the three curves correspond to the $u_{PP}$ values in the legend of panel (f). (h) Same as in (g) but for $\gamma=55^{\degree}$. 
} 
\label{fig:fig1}
\end{figure*}

We consider a dielectric sphere containing three point charges, a negative one positioned at the center of the sphere and two positive ones, equally charged and symmetrically placed at a distance $a$ from the center. This distribution of charges gives rise to a linear, axially symmetric quadrupole. The resulting electrostatic pair interaction is thus anisotropic and, given a set of microscopic parameters, it can be explicitly computed under linear approximation within a mean field approach~\cite{Bianchi2011Inverse,bianchi:2015}.  
We refer to this potential as ``DLVO-like" as in the limit of a single, central charge such a mean field interaction coincides with the well-known DLVO potential between homogeneously charged spheres.
The Inverse Patchy Particle (IPP) model discussed in the following represents the coarse-grained version of the aforementioned DLVO-like potential and as such can be regarded as representative of the effective interactions in heterogeneously charged systems such as globular proteins and patchy colloids~\cite{Bianchi2011Inverse,bianchi:2015}.

Within the IPP model, each particle has radius $\sigma_c=0.5>a$, which sets the units of length, and is endowed with three interaction sites, positioned exactly as the three charges of the mean field description. Each interaction site is the center of an interaction sphere. The off-center spheres emerge from the surface of the central sphere, thus defining the polar patches and the complementary equatorial belt, that is the part of the particle surface not covered by the patches. This geometry mimics the heterogeneous pattern of the surface charge distribution of the mean field model: the equatorial regions of two different IPPs as well as two patches of two IPPs  mutually repel each other, while a patch of one IPP is attracted to the equatorial region of a different IPPs. This consideration also explains the use of the ``inverse" patchy particles notion: unlike conventional patchy systems, the patches of IPPs can not bond to each other but rather repel each other.

The sphere associated to the central site has radius $\sigma_c+\delta/2$, while the off-center spheres have radius $\sigma_p$ constrained by $\sigma_p + a = \sigma_c + \delta/2$. The above constraint, which is a direct consequence of the screening conditions of the solvent, forces the off-center spheres to extend exactly up to the extension of the central sphere, i.e., $\delta$ is the sole parameter characterizing the interaction range of the model: if the center-to-center distance between two IPPs is $r$, then
$r<2\sigma_c$ implies an infinite, hard-sphere repulsion, while $r> 2\sigma_c+\delta$ implies that the two particles do not interact at all. The geometry of an IPP is hence specified by two parameters, $\delta$ and $a$, given the constraint on $\sigma_p$. An alternative way to characterize the model is to replace $a$ with the semi-opening angle of the off-center spheres, $\gamma$, which quantifies the surface area covered by a patch. The constraint on $\sigma_c$ translates in the expression $\gamma=\arccos{[(\sigma_c^2 + a^2 - \sigma_p^2) / 2a\sigma_c]}$. See Figure~\ref{fig:fig1} (panels a and b) for a detailed representation of the geometric parameters of the model.

As our coarse-grained description aims at accurately reproducing the DLVO-like description while being computationally efficient, the distance- and orientation-dependent pair interaction energy is written in the form~\cite{Bianchi2011Inverse,bianchi:2015}
\begin{equation}\label{eq:potential}
U (r,\Omega)  = \sum_{\alpha\beta} \epsilon_{\alpha\beta}w_{\alpha\beta}(r,\Omega) \, .
\end{equation}
In the above expression, $r$ is the center-to-center distance between two IPPs and $\Omega$ is their mutual orientation, $\alpha$ and $\beta$ identify the three interaction sites of the two particles, i.e., they run over the central site and both the off-center sites, $\epsilon_{\alpha \beta}$ characterizes the energy strength of the $\alpha \beta$ interaction, i.e., the interaction strength between the $\alpha$ site of one particle and the $\beta$ site of the other particle, and finally $w_{\alpha\beta}$ takes into account the interaction geometry of the specific pair configuration. While the values of $\epsilon_{\alpha \beta}$ are constant and characterize a specific set of microscopic parameters, the functions $w_{\alpha\beta}$ characterize the dependence on the mutual orientation and distance of the specific pair configuration. Such a dependence is chosen to be represented by the relative overlap volume between the interaction spheres associated to the interactions sites $\alpha$ and $\beta$ of the two particles~\cite{Bianchi2011Inverse,bianchi:2015}.
From an operational point of view, given a distance $2\sigma_c \leq r \leq 2\sigma_c+\delta$ and a relative orientation $\Omega$, the summation in Eq.~(\ref{eq:potential}) accounts for:
(i) the relative overlap between the spheres of radius $2\sigma_c +\delta$ associated to the two central sites, weighted by $\epsilon_{\text{center } \text{center}}$;  (ii) the relative overlap between the four spheres of radius $\sigma_p$ associated to the off-center sites of one particle and the two spheres associated to the central sites of the other particle, weighted by $\epsilon_{\text{center } \text{patch}}$; (iii) the relative overlap between the spheres associated to the off-center sites of the two different particles, weighted by $\epsilon_{\text{patch } \text{patch}}$; the relative overlap stands for the overlap volume between two spheres, normalized by the maximum possible overlap volume, i.e., the volume of the smallest sphere. 

The $\epsilon_{\alpha \beta}$ can be directly related to the charge balance between the different regions of the particle surface by a mapping between the IPP potential resulting
from Eq.~(\ref{eq:potential}) and the mean field, DLVO-like potential derived for a dielectric sphere with a given set of point charges~\cite{Bianchi2011Inverse,bianchi:2015,Notarmuzi_2024}. Here, however, instead of mapping the IPP model to specific parameter sets of
the DLVO-like description, we explore the role played by the net particle charge in a systematic fashion by varying the energy strengths arbitrarily. To this aim, we
fix the value of the interaction potential $U$ when the particles are at contact ($r=2\sigma_c$) and in one of three specific reference configurations, named $EE, EP, PP$ (Figure~\ref{fig:fig1} shows the three reference configurations in the legend of panels c and f): in the $EE$ configuration the symmetry axes of the particles are parallel, in the $EP$ configuration they are  orthogonal, and in the $PP$ configuration they are coincident. Once the desired energy strength of these configurations are defined and stored in a vector $\mathbf{u}=\{u_{EE}, u_{EP}, u_{PP}\}$, the vector $\boldsymbol{\epsilon} = \{\epsilon_{\text{center center}}, \epsilon_{\text{center } \text{patch}}, \epsilon_{\text{patch patch}}\}$ can be computed by solving
\begin{equation}\label{eq:system}
    W^{-1} \mathbf{u} = \boldsymbol{\epsilon}
\end{equation}
where $W^{-1}$ is the inverse of the matrix whose elements, $W_{\alpha \beta}^{AB}$, are the sum of all overlap volumes between all $\alpha\beta$ sites for the given AB configuration. The AB configurations are the reference configurations $EE, EP, PP$. 

In this work, we fix the interaction range to $\delta=0.2 \sigma_c$ and systematically vary the patch size and the net particle charge of the particles in order to assess the effect of the interplay between the geometry of the patches and the strength of the electrostatic
interactions on the liquid-liquid critical point. In particular, we vary $\gamma$ in the range $[30^{\degree}, 55^{\degree}]$ in steps of $5^{\degree}$, while we create a regular grid of values for $u_{EE}$ and $u_{PP}$, where $u_{EP}=-1.0$ sets the energy scale: we vary $u_{EE}$ and $u_{PP}$ independently in steps of $0.5$ within the range $[0, 2]$; we also add -- for all $\gamma$ values and selected $u_{EE}$ (namely, 0, 1 and 2) --  two values of $u_{PP}$ (namely, 4 and 6) to bridge towards IPPs with charge imbalances already studied in the literature~\cite{silvanonanoscale}. Note that a few data points (at large patch sizes and large $u_{EE}$ values) are missing due to the emergence of crystallization in the sample. 

\section{Methods}\label{sec:methods}

\subsection{Grand Canonical Monte Carlo simulations}

We perform Grand Canonical (GC) Monte Carlo (MC) simulations with a code adapted from the publicly available code published with Ref.~\cite{Rovigatti2018How}. Our code as well as the analytics tools used to produce the data presented in this paper are available at~\cite{GitCode}.
In a GCMC simulation the system energy $\cal E$ and the particle number $N$ are allowed to fluctuate so to estimate their probability distributions, while the volume of the cubic simulation box is fixed by its linear size $L=8$.   
A MC step corresponds to $N_{max}$ MC moves, where $N_{max}$ is the maximum number of particles allowed in the simulation box. The moves used in a MC step are the insertion/deletion of a particle, and a single particle rototranslational (RT) move, i.e., the contemporary  translation and rotation of a single particle~\cite{Rovigatti2018How}. An insertion/deletion move is attempted with probability 0.01, while a RT move with probability 0.99.
The maximum translation length (0.05) and maximum rotation angle (0.1) are chosen to result into an average acceptance rate of the RT move of about 30$\%$ around the critical point. The average acceptance rate strongly fluctuates between high values in the diluted phase and low values in the dense phase. 

\begin{figure*}[!ht]
\centering
 \includegraphics[width=0.9\textwidth]{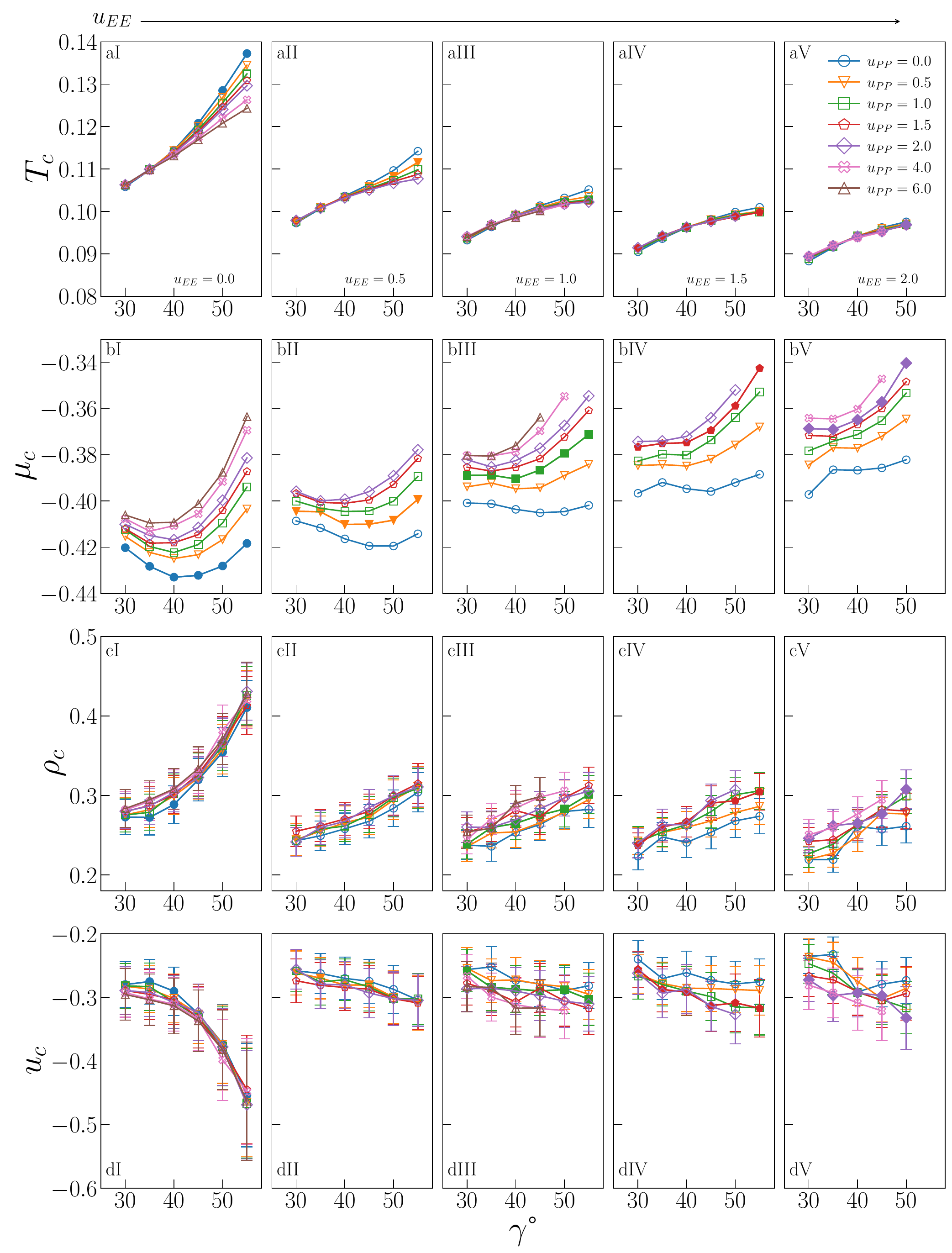}
\caption{Critical behaviour of all investigated IPP systems. From top to bottom: critical temperature $T_c$ (row a), critical chemical potential $\mu_c$  (row b), critical density $\rho_c$  (row c), critical energy density $u_c$  (row d). From left to right: values of $u_{EE}$ grow from $u_{EE}=0.0$ to $u_{EE}=2.0$ in steps of 0.5 per panel (labeled I to V). Different colors and symbols refer to different values of $u_{PP}$ as reported in the legend of panel (aIV).
} 
\label{fig:crit_params}
\end{figure*}

\subsection{Identification of the critical point}

For each model, we first perform a large number of short simulations at different values of the temperature $T$ and of the chemical potential $\mu$, so to approximately locate the phase separation region. We then select a few  values of $T$ and $\mu$ and perform 12 independent GCMC simulations per state point. Each simulation begins with $N_0=180$ particles and equilibrates for $2.5 \cdot 10^6$  MC steps, a value that is a posteriori checked to guarantee a sufficiently large equilibration time for all systems. The total run time per simulation is set to $5 \cdot 10^7$ MC steps, during which the values of $N$ and $\cal E$ are collected every $10^3$ MC steps and one
configuration is saved every $5\cdot10^4$ MC steps, thus resulting in a total of $57\cdot10^4$ values of $\cal E$ and $N$ per state point and 11400 configurations. 

At each state point, we calculate the scaling variable $\mathcal{M}=N+s\cal E$, where $s$ is a fitting parameter with non-universal values. As, at the critical point, the probability distribution of $\mathcal{M}$ coincides (up to vanishing second order corrections) with the distribution of the magnetization of the Ising model~\cite{Bruce1992Scaling}, the histograms produced by a simulation of the state point $(T,\mu)$ are reweighted~\cite{Ferrenberg1988New}, so to identify new values of $(T', \mu')$ and an optimal value of $s$ such that the distribution of $\mathcal{M}$, rescaled to have unit variance, matches the Ising magnetization distribution, computed as in Ref.~\cite{Tsypin2000Probability}. We perform simulations until the norm of the difference between the reweighted distribution of $\mathcal{M}$ and the Ising magnetization distribution is lower than 0.140. The final values of $T'$ and $\mu'$ are then defined to be the critical ones, $T_c$ and $\mu_c$.
We then define the critical density, $\rho_c$, and the critical energy density, $\epsilon_c$, as the average of their respective distributions at $(T_c, \mu_c)$, computed via histogram reweighting. 
After the identification of $(T_c, \mu_c)$, for some selected systems we also perform simulations at the critical point, in order to gather data for the structural properties of the critical phases. 
To verify whether a simulation is sufficiently close to the critical point we check that the distribution of $\mathcal{M}$ coincides with the Ising magnetization distribution without any reweighting.
The structural properties of the models characterized by $(u_{EE},u_{PP})=(0.0, 0.0)$ and
$(u_{EE},u_{PP})=(0.5,2.0)$ are computed by using configurations sampled at the critical point. 

\section{Results}\label{sec:results}

\subsection{The critical point}\label{sec:criticalpoints}

The behaviour of $T_c$, $\mu_c$, $\rho_c$ and $u_c$ is shown in Figure~\ref{fig:crit_params} for all investigated IPP systems.

The critical temperature (a-panels of Figure~\ref{fig:crit_params}) monotonically grows with $\gamma$ for any combination $(u_{EE},u_{PP})$ of the electrostatic repulsion, where $u_{EE}$ and $u_{PP}$ have nonetheless a different quantitative impact on $T_c$. The repulsion between the equators has in fact the strongest effect: on increasing $u_{EE}$ (from panel aI to aV of Figure~\ref{fig:crit_params}), $T_c$ significantly decreases for each given $\gamma$. It must be noted that, as $u_{EE}$ increases, the growth of $T_c$ with $\gamma$ becomes less and less pronounced at any fixed $u_{PP}$ and we observe a change in the curvature in the $\gamma$-dependence of $T_c$ from convex to concave. In contrast, the repulsion between the polar regions plays a significant role only when the $EE$ repulsion is small and, even in that case, only at large $\gamma$s (panels aI-aIII of Figure~\ref{fig:crit_params}). On increasing $u_{EE}$, the effect of $u_{PP}$ at large $\gamma$s reduces until it becomes negligible (panels aIV and aV of Figure~\ref{fig:crit_params}). Overall, the interplay between geometry and electrostatics leads to strong variations in $T_c$, from a minimal value 0.0883 to a maximal 0.1372, a value that is 55\% larger than the minimum.

The critical chemical potential (b-panels of Figure~\ref{fig:crit_params}) shows qualitatively different trends. In particular, $\mu_c$ displays a more pronounced dependence on the $PP$ repulsion: a growth of $u_{PP}$ implies an increase of $\mu_c$, meaning that all curves are shifted upward, regardless of $u_{EE}$ and $\gamma$.
The dependence on $u_{EE}$ is also monotonic, with $\mu_c$ growing on increasing $u_{EE}$, where the $EE$ repulsion has nonetheless a smaller effect with respect to the $PP$ repulsion. In contrast to $T_c$, $\mu_c$ does not grow monotonically with $\gamma$ at all values of the $EE$ repulsion: while for large values of $u_{EE}$, $\mu_c$ monotonically grows with $\gamma$ (see e.g. panel bV of Figure~\ref{fig:crit_params}), as $u_{EE}$ diminishes, $\mu_c$ shows instead a non-monotonic $\gamma$-dependence (see e.g. panels bI of Figure~\ref{fig:crit_params}).
A minimum of $\mu_c$ at an intermediate $\gamma$ implies that inserting in the system a particle with a smaller or larger patch is more costly. In the purely attractive case (i.e. $u_{EE}=u_{PP}=0$) the curve is almost symmetric with respect to its minimum, suggesting that particles with intermediate $\gamma$s can be inserted at a lower cost due to geometric reasons. On increasing only the $PP$ repulsion, the minimum does not move in $\gamma$, but it becomes increasingly costly to insert a particle with a large patch compared to one with a small patch, confirming that it is the number of unfavorable configurations due to the $PP$ repulsion to determine $\mu_c$: the larger and more repulsive the patches are, the more costly it is to insert the particles on average. When $u_{EE}$ also increases, $\mu_c$ gradually returns to being monotonic, as the $EE$ repulsion outweighs the $PP$ repulsion. The distinct behavior of the chemical potential compared to the other critical parameters and fields is thus an effect of the increased sensitivity of $\mu_c$ to the $PP$ repulsion.

Similar to $T_c$, the critical density (c-panels of Figure~\ref{fig:crit_params}) is a monotononically growing function of $\gamma$ for any combination $(u_{EE},u_{PP})$ of the electrostatic repulsion. Again like $T_c$, the growth is more pronounced when $u_{EE}$ is small and it flattens as the $EE$ repulsion increases, while the growth of $u_{PP}$ weakly affects $\rho_c$, regardless of $u_{EE}$ or $\gamma$. The remarkable changes in $\rho_c$ caused by the interplay between electrostatic repulsion and patch geometry imply that the largest value of the critical density, 0.430 (obtained for $u_{EE}=0.0$, $u_{PP}=2.0$ and $\gamma=55^{\degree}$), is 98\% larger than the smallest value, 0.219 (obtained for $u_{EE}=2.0$, $u_{PP}=0.0$ and $\gamma=35^{\degree}$).

The critical energy (d-panels of Figure~\ref{fig:crit_params}) mirrors the behaviour of $\rho_c$, which comes as no surprise given the strong correlation between these variables at the critical point~\cite{Wilding1995Liquid}. For $u_{EE}=0.0$, $u_c$ rapidly decreases as $\gamma$ grows, reaching a minimum value -0.469 for $u_{EE}=0.0$, $u_{PP}=2.0$ and $\gamma=55^{\degree}$ (the system with the largest $\rho_c$). The maximum value is -0.232 for $u_{EE}=2.0$, $u_{PP}=0.0$ and $\gamma=35^{\degree}$ (the system with the smallest $\rho_c$). Again in analogy to the critical density, $u_c$ is weakly affected by the $PP$ repulsion at any $\gamma$ and $u_{EE}$, while it increases with the $EE$ repulsion at any $\gamma$.

\begin{figure*}[!h]
\centering
  \includegraphics[width=\textwidth]{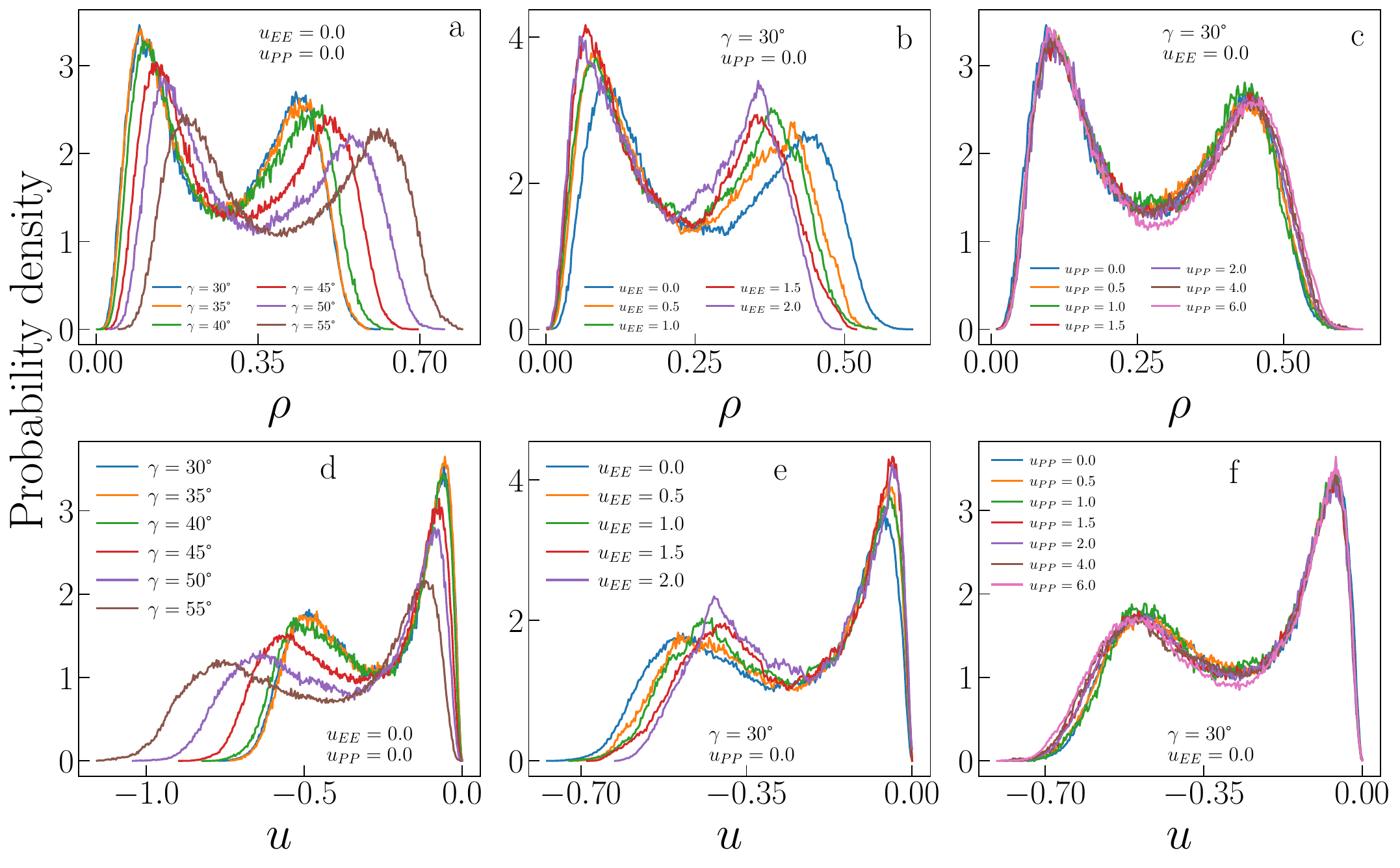}
\caption{Critical distributions of sample IPP systems obtained using histogram reweighting.
(a) Critical distributions of the density $\rho$ for $u_{EE}=u_{PP}=0.0$ and different values of $\gamma$.
(b) Critical distributions of $\rho$ for $\gamma=30$°, $u_{PP}=0.0$ and different values of $u_{EE}$.
(c) Critical distributions of $\rho$ for $\gamma=30$°, $u_{EE}=0.0$ and different values of $u_{PP}$.
(d), (e), (f) as in (a), (b), (c) respectively but for the critical distributions of the energy density $u$.
} 
\label{fig:crit_distrs}
\end{figure*}

Figure~\ref{fig:crit_distrs} allows to better understand the behaviour of $\rho_c$ and $u_c$ by displaying the critical distributions of these variables, as obtained after histogram reweighting. For small values of $u_{EE}$ and $u_{PP}$ (for instance for $u_{EE}=u_{PP}=0$), the distributions become wider and wider on increasing $\gamma$, with the weight of extremely large densities growing systematically with the patch size, subtracting weight to regions of low density (panel a of Figure~~\ref{fig:crit_distrs}). This behaviour is mirrored by the growth of the weight of very low energy regions and by the contemporary reduction of weight associated to regions of relatively high energy (panel d of Figure~\ref{fig:crit_distrs}). In contrast, when the opening angle and the PP repulsion are kept fixed and small, but the EE repulsion grows, we observe the opposite trend (panels b and e of Figure~\ref{fig:crit_distrs}). In this case, the weight moves toward regions of low density (high energy) with increasing $u_{EE}$, implying a decrease (growth) of $\rho_c$ ($u_c$). Finally, variations of $u_{PP}$ are substantially ineffective in altering the probability distributions of $\rho$ and $u$ for a given set of $\gamma$ and $u_{EE}$ values (panels c and f of Figure~\ref{fig:crit_distrs}) and indeed their average weakly depends on the $PP$ repulsion.

\begin{figure*}[!h]
\centering
  \includegraphics[width=\textwidth]{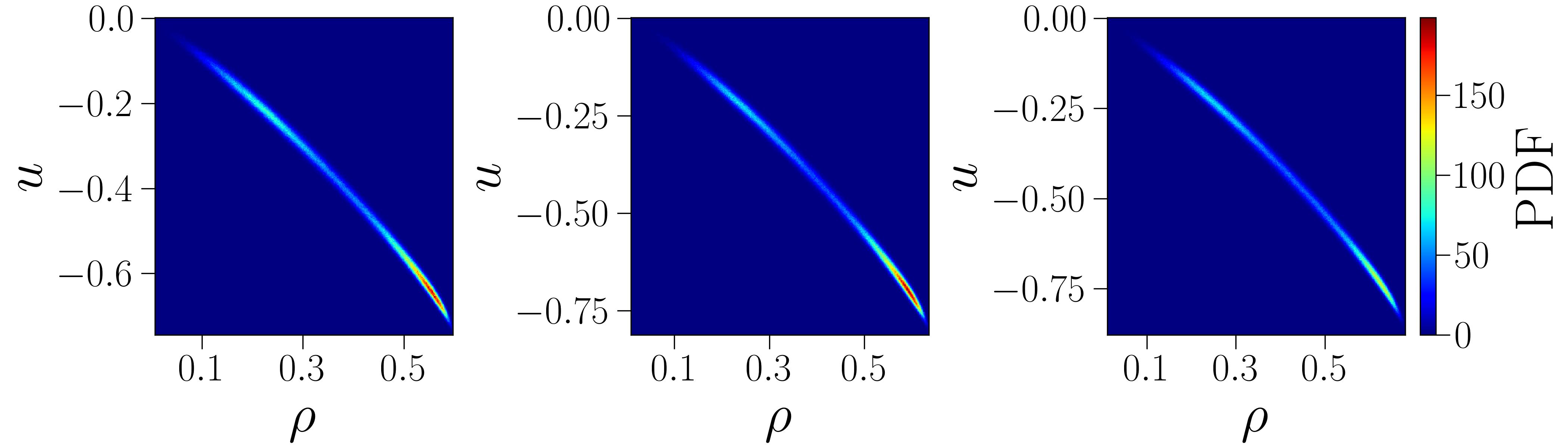}
\caption{Joint probability density functions of energy density $u$ and particle density $\rho$ at the critical point for $u_{EE}=u_{PP}=0.0$ and three different values of $\gamma=30^{\degree}, \, 40^{\degree}, \, 50^{\degree}$ from left to right.
} 
\label{fig:PNE}
\end{figure*}

As a further confirmation that the behavior of the critical energy can be explained by the strong correlation between $u_c$ and $\rho_c$ at the critical point, we show in Figure~\ref{fig:PNE} the joint probability density function of particle and energy density, computed from simulations at the critical point. The shape of the distributions confirm the existence of a strong correlation between these variables: regardless of $\gamma$ the distributions are extremely narrow along a slightly curved line so that the value of one variable is almost entirely determined by the value of the other, with extremely small fluctuations around the conditioned average. Moreover, the distributions clearly show that low values of $u$ are systematically associated to large values of $\rho$ and vice versa, hence confirming that the two opposite monotonic trends (with $\gamma$ and on increasing $u_{EE}$) shown in Figure~\ref{fig:crit_params} are due to the strong correlation between the critical fields. 


\subsection{Bonding volume $versus$ critical temperature}\label{sec:VbTc}

A well established result in the study of the critical point of patchy systems is that the critical temperature can be related to the amount of physical space around a single particle that is available for bonding. This quantity is referred to as bonding volume $V_b$ and can be defined as~\cite{Sciortino2007Self}
\begin{equation}\label{eq:Vb}
    V_b = \int \Theta[-U(\mathbf{r}, \Omega_1, \Omega_2)]  \frac{d\Omega_1 d\Omega_2}{16 \pi^2} d\mathbf{r}\, 
\end{equation}
where $\Theta$ is the Heavyside stepfunction, $r$ is the distance between particle 1 and 2, and $\Omega_{1}$ ($\Omega_{2}$) is the random orientation of particle $1$ ($2$). The above integral can be estimated by first performing the integrals with respect to $d\Omega_1 d\Omega_2$: this operation provides an estimate of the amount of physical space one particle has available to bonding when the other particle is at a distance $r$; subsequently, the integration over $\mathbf{r}$ can be performed~\cite{Sciortino2007Self, Wertheim1986Fluids}, leading to the value of $V_b$ for the selected set of model parameters.  The same result can be obtained by assigning to one particle
of the pair a random position (within the interaction range) and a random orientation. In this way, the whole integral in Eq.~\ref{eq:Vb} can be estimated with a single sampling as
\begin{equation}\label{eq:Vb_numeric}
    V_b = 4 \pi (r_{max}-r_{min}) \sum_{i=1}^Z \frac{r_{i}^2}{Z} \Theta[-U(\mathbf{r}_i, \Omega_{1i}, \Omega_{2i} )] 
\end{equation}
where $r_{max} - r_{min}=\delta$ is the interaction range, $Z$ is the number of configurations sampled; crucially, $r_i$ must be uniformly sampled from the surface of a sphere. Note that this procedure is completely general and can be used to estimate any integral of the form of Eq.~(\ref{eq:Vb}) for a generic function $F(\mathbf{r}, \Omega_1, \Omega_2)$ by simply replacing 
$\Theta$ with $F$ in Eq.~(\ref{eq:Vb_numeric}). 

\begin{figure*}[h]
\centering
  \includegraphics[width=\textwidth]{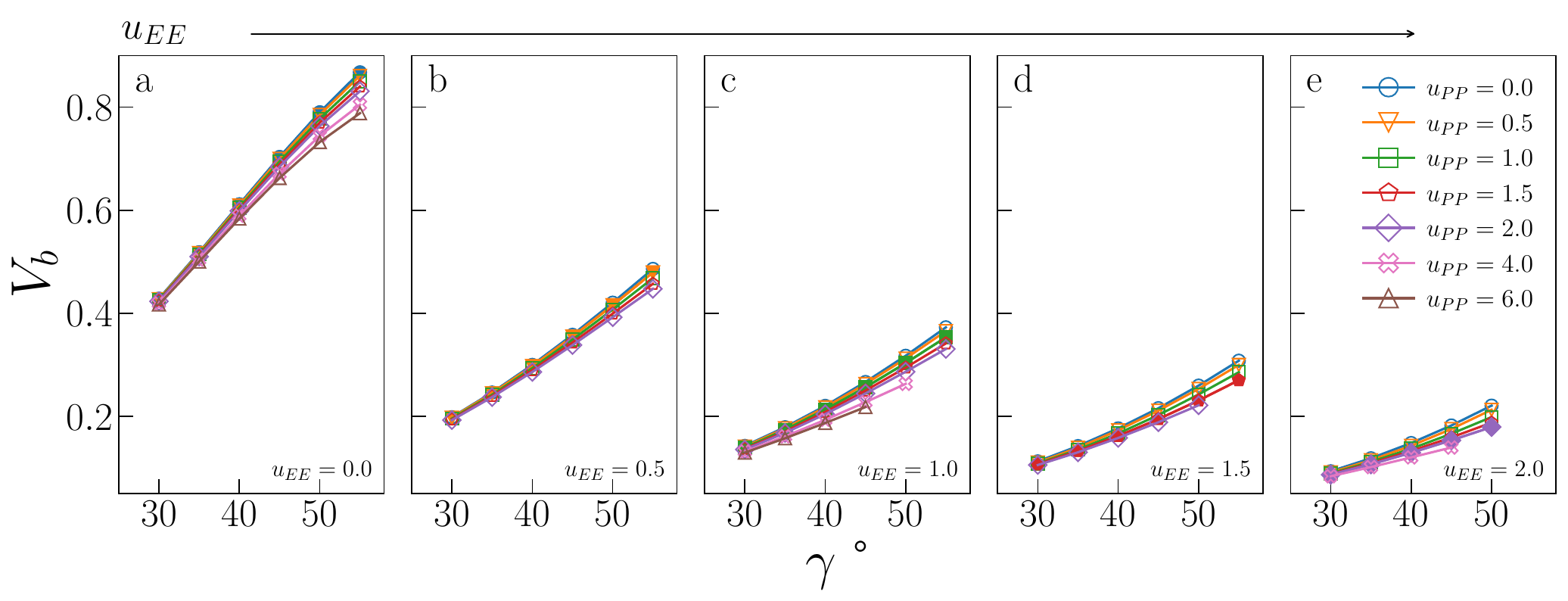}
\caption{Bonding volume $V_b$ for all systems studied in Figure~\ref{fig:crit_params}. The value of $u_{EE}$ grows from $u_{EE}=0.0$ to $u_{EE}=2.0$ in steps of 0.5 per panel (from a to e). Different colors and symbols refer to different values of $u_{PP}$, as shown in the legend of panel (e). Markers are filled when $u_{EE}=u_{PP}$ and empty otherwise. 
} 
\label{fig:V_b}
\end{figure*}

Figure~\ref{fig:V_b} displays $V_b$ for the systems considered in Figure~\ref{fig:crit_params}. As expected, the behaviour of $V_b$ reproduces the same trends observed for $T_c$: it strongly decreases on increasing $u_{EE}$ (from panel a to e) and it is always a monotonically growing function of $\gamma$; furthermore, it is weakly affected by $u_{PP}$. In contrast to $T_c$, the curvature of $V_b$ goes from concave to convex on increasing $u_{EE}$. Moreover, again in contrast to $T_c$, $V_b$ still decreases on increasing $u_{PP}$ at large values of $u_{EE}$ and $\gamma$. 

It is worth noting that, while in conventional patchy colloids, $V_b$ is in a straightforward relation with the number and size of the attractive patches~\cite{KernFrenkel_2003,Bianchi2006Phase}, in IPP systems $V_b$ emerges as a consequence of the interplay between electrostatics and geometry, which both contribute to control the particle bonding valence. As complex as this interplay may be, $V_b$ represents a powerful tool to estimate the critical temperature behaviour of sets of IPPs, since it is a thermodynamic-independent parameter based on pair properties.

\subsection{Second virial coefficient and effective particle's valence}\label{sec:b2}

The second virial coefficient, $b_2(T)$, is defined as~\cite{Noro2000Extended}
\begin{equation}\label{eq:b2}
    b_2(T) = -\frac{1}{2}\int [\exp(-\beta U(\mathbf{r}, \Omega_1, \Omega_2) - 1]  \frac{d\Omega_1 d\Omega_2}{16 \pi^2} d\mathbf{r}
\end{equation}
and quantifies the contribution of the pair-wise interaction to the equation of state of an ideal gas. The reduced second virial coefficient, $b_2^*(T)$, has been proposed by Noro and Frenkel as a scaling variable to extend van der Waals law of corresponding states to systems with variable attraction range~\cite{Noro2000Extended} and, since then, it has been used to map phase diagrams of different models in a large variety of systems, from proteins~\cite{Egelhaaf_2015} to colloids~\cite{ciulla}. The reduced second virial coefficient is defined as
\begin{equation}
    b_2^*(T) = b_2(T)/\left( 2 \pi \sigma_{eff}^3/3\right)
\end{equation}
where $2 \pi \sigma_{eff}^3/3$ is the second virial coefficient of a system of hard spheres with diameter $\sigma_{eff}$ and  $\sigma_{eff}$ can be calculated as
\begin{equation}\label{eq:sigma_eff}
    \sigma_{eff} = \int_0^{\infty} [1 - \exp(-\beta U_{rep}(r)]  dr  \, ,
\end{equation}
where $U_{rep}(r)$ is the repulsive part of the potential, i.e., $U(r) \Theta[U(r)]$. Defined in this way, $\sigma_{eff}$ quantifies the extension of the repulsive region of a generic potential in terms of a system of equivalent hard spheres
by weighting, in a temperature-dependent fashion, the strength of the repulsion between particles.

A generalized law of corresponding states for conventional patchy systems has been proposed under the observation that systems with the same number of patches tend to display similar values of $b_2^*$ at the critical temperature~\cite{Foffi2007}. 
A natural question to ask is whether the generalized law of corresponding states holds for IPP systems. To answer this question, we must calculate $b_2$ and $\sigma_{eff}$. For the first, one can rely on the observation that Eq.~(\ref{eq:b2}) has the same form of Eq.~(\ref{eq:Vb}), thus, if the Heavyside function is replaced by the Meyer function $[1 - \exp(-\beta U_{rep})]$, we can use Eq.~(\ref{eq:Vb_numeric}) for measuring $b_2$. Eq.~(\ref{eq:sigma_eff}), however, does not have the same form of Eq.~(\ref{eq:Vb}) and hence a different strategy is required. The difficulty arises in front of the observation that the potential of IPP systems has a repulsive component that is not radial but rather depends on the relative orientation between the particles and hence the radial integral in Eq.~(\ref{eq:sigma_eff}) is not appropriate to quantify $\sigma_{eff}$: an evaluation of $U_{rep}$ necessarily requires the exploration of the whole configuration space, i.e., an integration with respect to  $d\Omega_1 d\Omega_2 d\mathbf{r}$, with the consequence that the resulting integral has the dimension of a volume. Simply replacing the integration with respect to $r$ in Eq.~(\ref{eq:sigma_eff}) with one over the configuration space and then taking the cubic root of the resulting integral is clearly inappropriate, as for isotropic potentials it would not yield the same result of Eq.~(\ref{eq:sigma_eff}). Hence, we generalize the definition of $\sigma_{eff}$ as
\begin{equation}\label{eq:sigma_eff_new}
    \sigma_{eff} = \int_0^{\infty} \frac{1 - \exp[-\beta U_{rep}(\mathbf{r}, \Omega_1, \Omega_2)]}{4 \pi r^2}  \frac{d\Omega_1 d\Omega_2}{16 \pi^2} d\mathbf{r}. 
\end{equation}
Note that Eq.~(\ref{eq:sigma_eff_new}) reproduces exactly the same results as of  Eq.~(\ref{eq:sigma_eff}) for isotropic potentials as well as for conventional patchy systems. The integral in Eq.~(\ref{eq:sigma_eff_new}) is evaluated using again
Eq.~(\ref{eq:Vb_numeric}) where the Heavyside stepfunction is replaced by the function $[1 - \exp(-\beta U_{rep})] / (4 \pi r^2)$.

The comparison of the second virial coefficients of different IPP systems, however, is not straightforward even once the measures of $b_2$ and $\sigma_{eff}$ are well-defined. The observation made in Ref.~\cite{Foffi2007}
relates the second virial coefficient of different patchy systems to the number of patches per particles. Under the single bond per patch condition, such a quantity corresponds to the maximum number of energetic bonds per particle~\cite{Foffi2007}, often referred to as particle functionality. As in IPP systems the particle functionality is not a built-in feature of the model, we need to determine the maximum number of bonds that an IPP can in principle form. For the purpose of our discussion in subsections~\ref{sec:PProle} and~\ref{sec:GbEb} we distinguish between geometric and energetic bonds: a geometric bond, $G_b$, forms between two particles when their distance $r$ is $2\sigma \leq r \leq 2\sigma + \delta$; an energetic bond, $E_b$, is a geometric bond with pair energy $U<0$. Note that in conventional patchy systems all bonds are energetic bonds. While the maximum number of geometric bonds that an IPP can form is 12, the so called kissing number in three dimension,  the maximum number of energetic bonds is the particle functionality, which we thus label $f_E^{max}$. To infer $f_E^{max}$, we devise a specific MC sampling with 12 particles positioned around a central particle along the vertices of a regular icosahedron, where each of the external particles is within a distance $r < 2\sigma_c+\delta$ from the central one. These 12 particles are roto-translated by selecting one at random and moving its center of mass by a vector with three different random components between $-\Delta/2$ and $\Delta/2$. The move is accepted if the particle remains within the interaction range of the central particle,  and if no overlap is created, nor with the central particle nor with the 11 remaining external  particles. Basically, any move that keeps the number of geometric bonds equal to 12 without creating overlaps is accepted. $\Delta=0.17$ is selected so to have an average acceptance rate of the move around 30\%. In this way, we create a large number of random configurations where the central particle has 12 possibly bonded neighbours. Note that this first stage of the MC is not concerned with the specific interaction potential of IPP systems and can simply be seen as a way to sample configurations where the kissing number of the central sphere is 12 given a square well potential with interaction range $\delta$. In particular, the value $\Delta=0.17$ is independent on the interaction potential of the IPP model for which the measure is being performed.   In the second stage of the MC, the IPP potential is activated and the measure of $f_E^{max}$ is performed. In this second stage, external particles are first moved in the same exact way as they were moved in the first stage, so to have again an average acceptance rate around 30\%. Moves are accepted according to the same criterion described above. If a move is accepted, a random orientation is assigned to the newly positioned particle and the  IPP potential between it and the central particle is computed, so to evaluate if the new position and orientation of the moved particle forms an energetic bond with it. This scheme allows to monitor the evolution of the total number of energetic bonds formed by the central particle as the remaining 12 are roto-translated (with respect to it) and acquire a random orientation. $f_E^{max}$ is 
estimated as the largest number of bonds formed by the central particle along the simulation. In principle, this measure provides a lower bound on the quantity of interest, so we attempted to move a random particle for a total of $4 \cdot 10^{11}$ times, corresponding to have successfully moved each of the 12 external particles $10^{10}$ times each. This large number of measures makes us confident that our estimator of $f_E^{max}$  well captures the maximum number of energetic bonds a particle can form. Note, in particular,
that two successive configurations of the simulation are clearly correlated, but we are not interested
in the probability distribution of the number of energetic bonds formed by the central particle, rather our focus is in the maximum number of energetic bonds that has non-zero probability. Hence, there is no reason to disregard any sampled configuration because of correlations:  in principle, the measure becomes exact if the entire configuration space is sampled, even if the sampling is made of a series of strongly correlated configurations.

\begin{figure*}[h]
\centering
  \includegraphics[width=\textwidth]{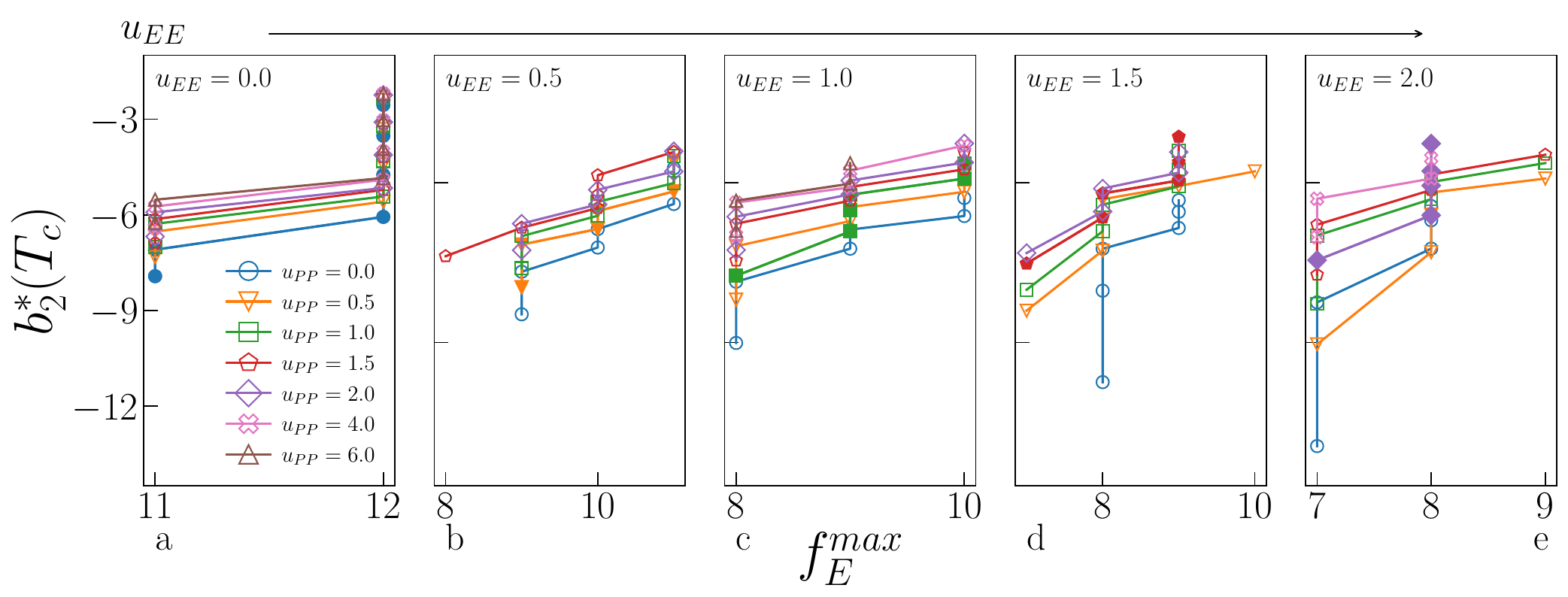}
\caption{Reduced second virial coefficient at the critical temperature, $b^*_2(T_c)$, as a function of the maximal functionality, $f_E^{max}$, for all systems studied in Figure~\ref{fig:crit_params}. Values of $u_{EE}$ grow from $u_{EE} = 0.0$ to $u_{EE} = 2.0$ in steps of 0.5 per panel (from a to e). Different colors and symbols refer to different values of $u_{PP}$ as shown in the legend of panel (a). Markers are filled when $u_{EE} = u_{PP}$ and empty otherwise. 
For each $u_{EE}$, $u_{PP}$ the smallest $b_2^*(T_c)$ corresponds to $\gamma$=30° and then
$\gamma$ grows by 5° pointwise, up to $\gamma=55$°, when $b_2^*(T_c)$ reaches its largest value. 
} 
\label{fig:b2_vs_fmax}
\end{figure*}

Figure~\ref{fig:b2_vs_fmax} shows the reduced second virial coefficient at the critical point as a function of $f_E^{max}$ for the IPP systems under investigation in this work. The behavior of the latter is clearly correlated with the bonding volume: $f_E^{max}$ diminishes as $u_{EE}$ grows (from panel a to e) and is rather insensitive to $u_{PP}$. Interestingly, the variability of $b_2^*(T_c)$ against variations of $u_{PP}$ depends on $u_{EE}$. At small EE repulsion the patch-patch interaction seems to weakly affect $b_2^*(T_c)$, while changes in $u_{PP}$ become more relevant in determining $b_2^*(T_c)$ for large equator-equator repulsion, meaning that the charge imbalance  strongly impacts the behavior of $b_2^*(T_c)$ for a given geometry. Despite the spectrum of values spanned by the $b_2^*(T_c)$ of IPP systems is consistent with the values observed for conventional patchy systems~\cite{Foffi2007}, it is not possible to apply the generalized law of corresponding states proposed by Foffi and Sciortino~\cite{Foffi2007}, meaning that it is not possible to identify classes of IPP systems with the same corresponding states $a$  $priori$ on the basis of their bonding functionality. 
However, the larger values of $b_2^*$, observed for small values of $u_{EE}$, are consistent with experimental measurements on monoclonal antibodies~\cite{Sibanda2023Relationship, Singh2019determination}, globular proteins~\cite{Zhang2012Charge, Egelhaaf_2015} and folded domains of intrinsically disordered proteins~\cite{Kim_2024}.

In the following we focus on the microscopic characterization of the aggregates at the critical point to better understand the behaviour of the described critical quantities.

\subsection{The role of the patch-patch repulsion}\label{sec:PProle}

The discussion above leads to the observation that the parameter $u_{PP}$ is the less relevant for the critical behaviour of our systems. The reason why this is the case is rooted in the ability of the particles to self organize into configurations where the $PP$ repulsion is completely avoided. To show this, we calculate the probability distribution of the interaction energy of random pairs of bonded particles and compare it to the probability distribution of the pair interaction energy measured in the simulations at the critical point, for several systems. 

\begin{figure*}[h]
\centering
  \includegraphics[width=\textwidth]{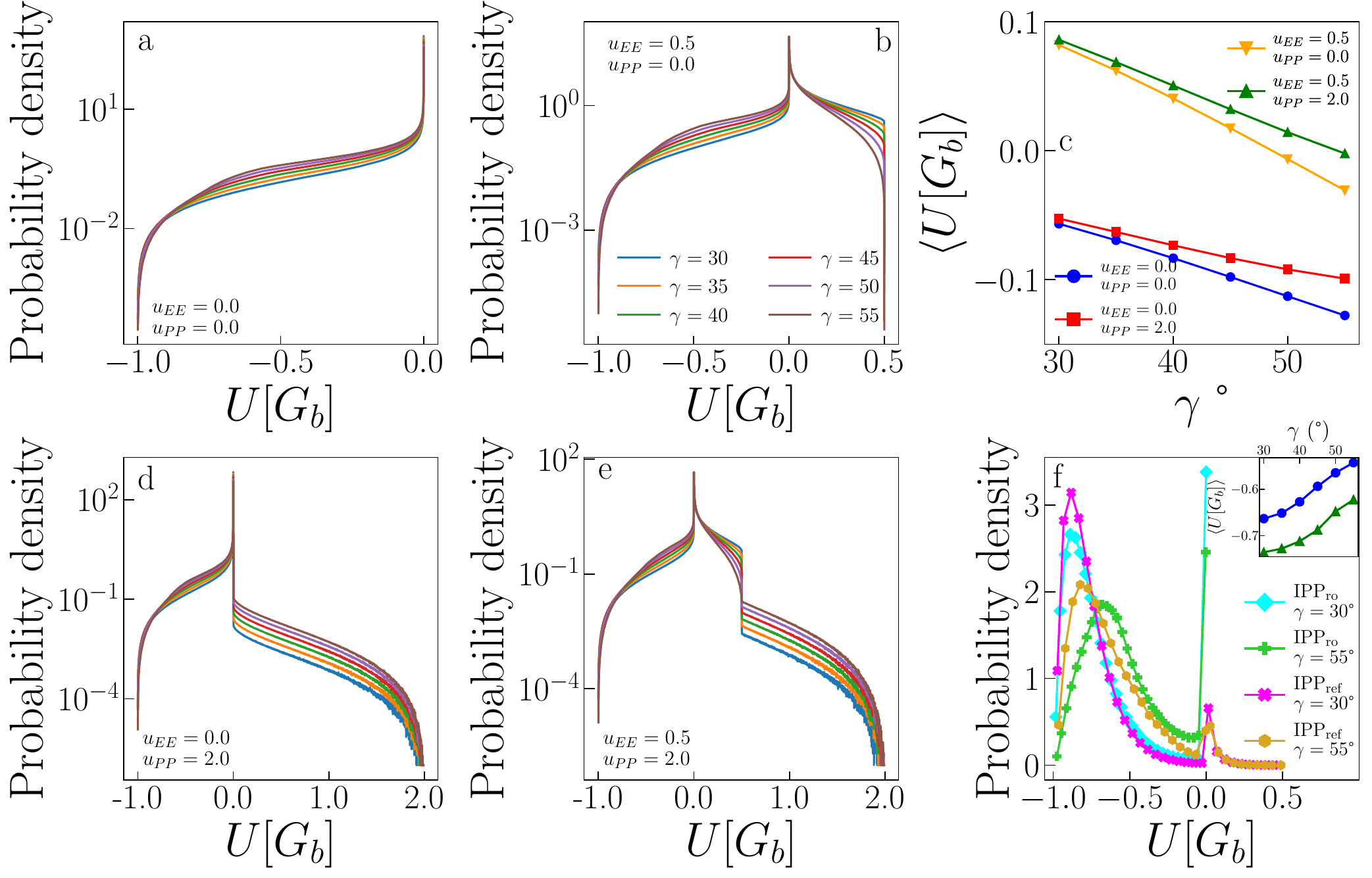}
\caption{Impact of patch-patch repulsion on the energy of IPP systems. (a) Distributions of the energy of geometric bonds for randomly generated pair configurations of IPPs with $u_{EE}=u_{PP}=0.0$ and different values of $\gamma$. (b) Same as in (a), but for IPPs with $u_{EE}=0.5$ and $u_{PP}=0.0$.
(d) Same as in (a), but for IPPs with $u_{EE}=0.0$ and $u_{PP}=2.0$. (e) Same as in (a), but for IPPs with $u_{EE}=0.5$ and $u_{PP}=2.0$. (c) Average of the distributions shown in panels a, b, d, e. (f) Distributions of the energy of geometric bonds for configurations observed in  simulations at the critical point for IPP$_{\rm{ro}}$ and IPP$_{\rm{ref}}$ systems, i.e., systems with $u_{EE}=0.0$, $u_{PP}=0.0$ and $u_{EE}=0.5$, $u_{PP}=2.0$ respectively, for $\gamma=30°$ and $\gamma=55°$. 
}
\label{fig:energy_distr}
\end{figure*}

Figure~\ref{fig:energy_distr} shows the probability distribution of the pair energy of geometric bonds for a variety of systems. Distributions in panels a, b, d, and e for the systems $(u_{EE}, u_{PP}) = (0.0, 0.0), \, (0.5, 0.0), \, (0.0, 2.0), \, (0.5, 2.0)$ (and all $\gamma$s), respectively, are computed by creating random geometric bonds. Specifically, while one IPP is fixed in position and orientation, the other is assigned a random position within the interaction distance and a random orientation. Panel c shows the average of these distribution as a function of $\gamma$, while panel f reports the distribution for the systems $(u_{EE}, u_{PP}) = (0.0, 0.0)$ and $(0.5, 2.0)$ at the two extreme values of $\gamma = 30^{\degree}$ and $55^{\degree}$. For consistency with Ref.~\cite{Notarmuzi_2024}, the system with $(u_{EE}, u_{PP}) = (0.0, 0.0)$ is referred to as IPP$_{\rm{ro}}$, where the subscript stands for ``repulsion off", and the system with $(u_{EE}, u_{PP}) = (0.5, 2.0)$ is named IPP$_{\rm{ref}}$, where the subscript stands for ``reference", as these values of the electrostatic repulsion have been observed in previously studied IPPs systems~\cite{silvanonanoscale,Bianchi2011Inverse,bianchi:2015,Noya2014}. 

The probability distributions of randomly generated configurations estimate the amount of possible pair configurations with a given energy. In absence of any electrostatic repulsion (panel a of Figure~\ref{fig:energy_distr}) the probability of a given energy is higher the greater (less negative) the value of the energy $U[G_b]$ is: while perfect $EP$ configurations (with $U[G_b]=-1$) are relatively rare, the distributions show a long regime of exponential growth at intermediate energies (with $-1<U[G_b]<0$, where the amplitude is higher for larger $\gamma$s) and a peak at $U[G_b]=0$. When either the $EE$ (panel b) or the $PP$ (panel e) repulsion is present, configurations with $U[G_b]>0$ become possible: the largest energies can be reached only if either the $EE$ or the $PP$ interaction, respectively, contributes to $U$. While configurations where the $EE$ repulsion contributes to the bond energy are relatively abundant (with $0<U[G_b]<0.5$), configurations where the $PP$ repulsion plays a role are exceedingly rare: as soon as $U[G_b] > 0$, the probability has a substantial drop, larger for small $\gamma$s, and then exponentially decays (note the log-lin scale) with a rate that seems $\gamma$-independent. This behaviour is confirmed also when both $u_{EE}\neq0$ and $u_{PP}\neq0$ (panel d), where the significant drop in the probability occurs as soon as $U[G_b] > u_{EE}$.

In summary, configurations that give a large $PP$ contributions to $U$ are not numerous, meaning that it is relatively simple for a pair of IPPs to avoid such configurations when forming a geometric bond in a simulation. This speculation is confirmed by the distributions shown in panel f: for IPP$_{\rm{ref}}$ systems the probability that $U[G_b]>0.5$ is very close to zero, thus explaining why $PP$ repulsion rarely impacts the critical parameters and fields. From  comparing IPP$_{\rm{ro}}$ and IPP$_{\rm{ref}}$ systems in panel f, we further note that the presence of electrostatic repulsion facilitates the formation of low energy bonds both at large and small $\gamma$s, and greatly reduces the probability of configurations with zero energy. The weight in the distributions of those configurations that have zero energy in IPP$_{\rm{ro}}$ is partially transferred to configurations with $U>0$, but a large fraction of this weight is actually moved to configurations with very low energy. 

As a result of the described differences between randomly generated pair configurations and pairs measured in simulations, the average energies of a geometric bond, $\langle U[G_b] \rangle$, (reported in panel c for random pairs and in the inset of panel f for pairs in simulations) show opposite trends: while $\langle U[G_b] \rangle$  decreases with $\gamma$ for random pairs, it instead increases with $\gamma$ in the simulations. 
In particular, panel c shows that the $EE$ repulsion has the largest effect on the average energy of a geometric bond: when this repulsion is off, then $\langle U[G_b] \rangle<0$, while a very mild $EE$ repulsion causes a shift of $\langle U[G_b] \rangle$ to higher and mostly positive values. In contrast, the $PP$ repulsion mainly tunes the rapidity with which the average energy decreases as $\gamma$ grows. 
In contrast, the inset of panel f shows that $\langle U[G_b] \rangle$ is always negative in simulations of both $\rm IPP_{ro}$ and $\rm IPP_{ref}$ and increases with $\gamma$: on increasing $\gamma$, bonds become thus weaker.  
Moreover at any fixed $\gamma$, the average energy of the system with only directional attraction is always higher than the average energy of the system with directional attraction and directional repulsion, this is due to additional morphological constraints introduced by the electrostatic repulsion leading to more optimized $EP$ configurations~\cite{Notarmuzi_2024}. 

\subsection{Geometric $versus$ energetic bonds}\label{sec:GbEb}

\begin{figure*}[h]
\centering
  \includegraphics[width=\textwidth]{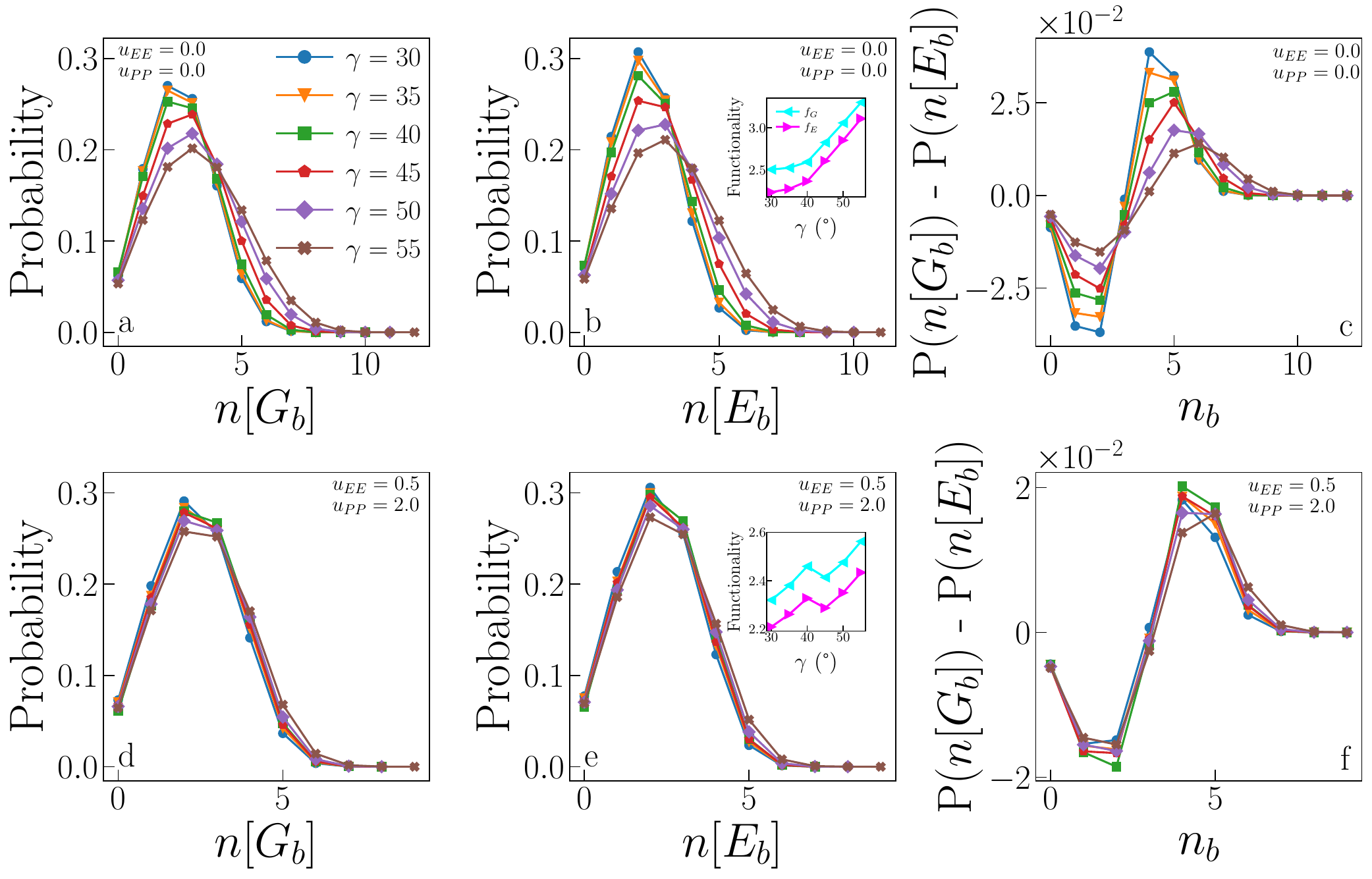}
\caption{Statistics of geometric and energetic bonds from samples collected at the critical point. (a) Probability that $n[G_b]$ geometric bonds are formed for systems with $u_{EE}=u_{PP}=0.0$ and different values of $\gamma$. (b) Probability that $n[E_b]$ energetic bonds are formed for the same systems as in panel (a). Inset: average functionality, i.e., average number of bonds (geometric and energetic, $f_G$ and $f_E$ respectively) as a function of $\gamma$, 
for the same systems as in panels (a) and (b). (c) Difference between the probability to form $n[G_b]$ geometric bonds and the probability to form $n[E_b]$ energetic bonds for the same systems as in panels
(a) and (b). (d), (e), (f) Same as in (a), (b), (c) respectively, but for systems with $u_{EE}=0.5$ and $u_{PP}=2.0$. 
}
\label{fig:bonds}
\end{figure*}

We now compare the
probability that a particle forms $n$ geometric bonds to the probability that a particle forms $n$ energetic bonds. Figure~\ref{fig:bonds} (panels a-b and d-e) displays these probabilities for IPP$_{\rm{ro}}$ (top) and IPP$_{\rm{ref}}$ (bottom) systems.  In absence of electrostatic repulsion (panels a and b), the distributions of both $n[G_b]$ and $n[E_b]$ show a remarkable dependence on $\gamma$, with large patches allowing more (geometric as well as energetic) bonds than small patches. Only a fraction of geometric bonds is also energetic: the probability of having a small number of energetic bonds is slightly higher than that of having the same number of geometric bonds, while for a large number of bonds the probability is higher that they are geometric rather than energetic (see the difference between the two cases reported in panel c). This trend is suppressed as $\gamma$ grows, when the two distributions become increasingly similar, suggesting that large patches allow for a greater ability to form energetic bonds. Note that, on increasing $\gamma$, the numerous energetic bonds formed tend to be weaker as shown in the inset of panel f of Figure~\ref{fig:energy_distr}.
The presence of electrostatic repulsion significantly alters the described scenario: the dependence on $\gamma$ is almost entirely suppressed, meaning that for large patches, the electrostatic repulsion acts against the formation of many energetic bonds. In other words, the fraction of geometric bonds which is also energetic does not significantly vary with $\gamma$ due to the electrostatic repulsion (see also panel f).

The insets of panels b and e of Figure~\ref{fig:bonds} show the average functionality of IPP$_{\rm{ro}}$ and IPP$_{\rm{ref}}$ systems at criticality.
In conventional patchy systems, the functionality $f$ of a particle is defined as the maximum number of bonds a particle can form and it corresponds to the number of patches per particle when the single bond per patch condition is satisfied. In IPP systems, however, the single bond per patch condition is not guaranteed and hence we define the functionality as the average number of bonds per particle actually formed in a simulations. As we distinguish between geometric and energetic bonds, we also distinguish between geometric and energetic functionalities, $f_G$ and $f_E$, respectively. Both quantities can be measured by averaging (over the whole system) the number of bonds each particle forms. Importantly, this definition implies that $f$ depends on the thermodynamic conditions the system is at. Data show that both functionalities grow with $\gamma$ for both systems, but in absence of repulsion (inset of panel b) the growth is such that the gap between $f_G$ and $f_E$ becomes smaller as $\gamma$ increases; this behaviour reflects the fact that the distributions of $n[G_b]$ and $n[E_b]$ become increasingly similar as $\gamma$ grows. In presence of electrostatic repulsion (inset of panel e) the average functionalities have a systematically smaller value if compared to the case where the electrostatic repulsion is absent, namely $2.2 \leq f_{G/E} \leq 3.4$ for $\rm IPP_{ro}$ systems and $2.2 \leq f_{G/E} \leq 2.6$ for $\rm IPP_{ro}$ systems; moreover,  the gap between $f_G$ and $f_E$ remains constant as the patch size grows. This is easily understood considering that identical configurations can have a much larger energy when $u_{EE}$ and/or $u_{PP}$ are non-zero: two repulsive equatorial regions of two IPPs surely interact as soon as the two particles are within their interaction distance, hence giving a positive contribution to the pair energy, regardless of $\gamma$.  That the reduced growth of the functionality with $\gamma$ is a consequence of mostly the $EE$ repulsion is an obvious consequence of the fact that configurations dominated  by patch-patch repulsion are avoided, as already discussed.  

It is worth noting that the average (geometric as well as energetic) functionality can be related to the compactness of the aggregates. Smaller functionalities imply more branched structures, as discussed in Ref.~\cite{Notarmuzi_2024}. In particular, $\rm IPP_{ro}$ systems with small patches have on average a small number of bonds, most of which are energetic, which give rise to branched structures, as observed in Ref.~\cite{Notarmuzi_2024}. These branched structures allow the system to condense into the liquid phase even at relatively low densities. In contrast, $\rm IPP_{ro}$ systems with large patches have on average a larger number of bonds, but a fraction of them is only geometric, which is due to the compact structures formed in absence of directional repulsion. This compact structures
require a large density for the liquid phase to condense, which explains the behaviour of $\rho_c$ with $\gamma$ in $\rm IPP_{ro}$ systems. In summary, low $f_{G/E}$ values imply branched structures, which in turn lead to low $\rho_c$ values. The same paradigm is observed for $\rm IPP_{ref}$ systems: as their $f_{G/E}$ values are systematically lower than those observed in $\rm IPP_{ro}$ systems and do not raise significantly with $\gamma$, their $\rho_c$ is also significantly lower over the whole $\gamma$-range, confirming that the electrostatic repulsion is a key factor in reducing the particle's connectivity. 

\section{Conclusions}\label{sec:conclusions}
In this work we numerically study the effect of electrostatic anisotropy on the LLPS of heterogeneously charged particles, referred to as IPPs, being them representative of charged patchy colloids or proteins systems. By taking advantage of a relatively simple coarse-grained model, we are able to investigate the critical behaviour of a large selection of IPP systems $via$ robust MC simulations. Our model reproduces the features of a directional screened Coulomb interaction for spherical particles with a simple charge heterogeneity and allows to control the competition between surface patchiness and charge imbalance by means of a few parameters. We stress that, despite our model is more suitable for colloids and globular rather than disordered proteins, estimates of the reduced second virial coefficient of our systems at the critical point are in the range reported not only for globular proteins but also for disordered proteins and antibodies. This supports the speculation that our modeling approach has a predictive power beyond the spherical approximation.

We show that anisotropic electrostatics results in a limited bonding valence, a feature that is usually associated only to site-specific interactions. In particular, we show that the directional attraction stemming from the interactions between oppositely-charged regions is not the only responsible for such a limited functionality: the directional repulsion stemming from like-charged regions is in fact crucial in controlling the bonding valence, thus implying that both charge patchiness and charge imbalance control the ability of a particle to form bonds. As an effect of the limited bonding functionality, the LLPS critical point shifts towards extremely low temperatures and densities. In particular, consistently with the LLPS behaviour of systems with site-specific interactions, the directional nature of the attractive interactions shifts the critical point towards lower temperatures and densities, where smaller patches disfavour the condensation of the dense liquid phase with respect to larger patches. Electrostatic directional repulsion further reduces the critical parameters, where the impact of the electrostatic repulsion on the critical point varies with the size of the patches, highlighting the complex interplay between charge imbalance and charge patchiness.

We rationalize the behaviour of the critical parameters in terms of thermodynamic-independent pair properties such as the particle bonding volume and the probabilities for a particle to form a given number of bonds or have a given energy. The collection of these quantities provides additional insight into the morphological features of the aggregates. In particular, while in systems with only directional attraction the number of possibly bonded pair configurations is controlled only by the patch size, when the directional repulsion is also present, the number of possibly bonded pair configurations is controlled by the complex interplay between patch size and charge imbalance. As a consequence, we observe the emergence of branched rather than compact structures not only at small patches -- as it is for IPPs with only directional attraction -- but also at large patches. This outcome highlights the potential of anisotropic electrostatics to control the LLPS by tuning the charge patchiness of the systems by means of, e.g., pH changes or, specifically for protein systems, mutagenesis.  

\section*{Author Contributions}
Both authors contributed equally to this work.

\section*{Conflicts of interest}
The authors declare no conflicts of interest.

\section*{Acknowledgements}
DN and EB acknowledge support from the Austrian Science Fund (FWF) under Proj. No. Y-1163-N27. Computation time at the Vienna Scientific Cluster (VSC) is also gratefully acknowledged. 

\renewcommand\refname{References}
\providecommand*{\mcitethebibliography}{\thebibliography}
\csname @ifundefined\endcsname{endmcitethebibliography}
{\let\endmcitethebibliography\endthebibliography}{}

\end{document}